\documentclass[]{spie}  
\usepackage{graphicx,subfigure}
\usepackage{amsmath,amsfonts,amssymb}
\usepackage{graphicx}
\usepackage[colorlinks=true, allcolors=blue]{hyperref}

\title{The structure monitoring of the MST prototype of CTA}

\author[a]{Victor Barbosa Martins}
\author[a]{Markus Garczarczyk for the The Cherenkov Telescope Array Consortium}
\affil[a]{DESY, Platanenallee 6, Zeuthen, Germany}

\authorinfo{Further author information: (Send correspondence to V.B.M)\\V.B.M.: E-mail: victor.barbosa.martins@desy.de}

\begin{document} 

\maketitle

\begin{abstract}
The Cherenkov Telescope Array (CTA) is the next generation of ground-based gamma-ray
observatory. The observatory will consist of two arrays, one located in the southern hemisphere (Paranal,Chile) and the
other in the northern hemisphere (Canary Island, Spain), covering the whole sky in the range of
observation. More than 100 telescopes are planned to be in operation for as long as 30 years, which
motivated the development of a continuous condition monitoring of the individual telescopes. The
main goal of the monitoring is to detect degradation and failures before critical damages occur. Two
approaches are considered: the structure monitoring system, in which the Eigenfrequencies of the
telescope and their damping rates are measured and monitored; and the drive monitoring, in which
the power spectra of rotating components are measured during telescope movements. The structure
monitoring concept system was applied to the prototype Medium Size telescope (MST) prototype of CTA in
Berlin during late 2018 and in 2019, and the first results are presented here. The system showed
reasonable stability during periods, in which the telescope structure was unchanged. The
system was also capable to detect mechanical changes, e.g. varying tension in the steel ropes of the
camera support structure. The successful implementation of the structure monitoring system supports the decision of implementing the system in all future MSTs.
\end{abstract}


\section{INTRODUCTION}
\label{sec:intro}  

The gamma-ray sky is investigated through satellite and ground-based detectors. While satellite detectors are mainly limited by their small collection area (ex.: FERMI-LAT \cite{fermilat} with 4x4 towers of area 40 $cm^2$ each), Imaging Atmospheric Cherenkov Telescopes (IACTs) provide a collection area of the order of some $km^2$ by means of an indirect detection. IACTs take advantage of stereoscopic detection of Cherenkov radiation emitted by the secondary particles in development of the atmospheric shower, which makes possible to reconstruct the direction and energy of the incoming photon. The current operating IACT facilities are the High Energy Stereoscopic System (H.E.S.S.)\cite{hessicrc} in Khomas Highland (Namibia) composed by 5 telescopes,  the Major Atmospheric Gamma Imaging Cherenkov Telescope (MAGIC)\cite{magic} in La Palma (Spain) composed by 2 telescopes and the Very Energetic Radiation Imaging Telescope Array System (VERITAS)\cite{veritas} composed by 4 telescopes. Those telescopes are in operation since the years 2000s and have already discovered a number of gamma-ray sources. Better spatial and time resolutions and a broader energy range are necessary to investigate further the deepest and most interesting science cases in gamma-ray astronomy \cite{sciencewithcta}. 

The Cherenkov Telescope Array (CTA) is the next generation of ground-based gamma-ray observatory and it will provide an energy coverage from 20 GeV to 300 TeV, a sensitivity level improvement of an order of magnitude at 1 TeV in comparison to current instruments and the full sky coverage, with sites in the two hemispheres (Cerro Paranal in Chile and La-Palma in Spain). Three kind of telescopes are planned to be used in the observatory to provide this energy coverage: the small-sized telescope (SST), the medium-sized telescope (MST) and the large-sized telescope (LST). The SSTs are better suited for the highest energies, while the LSTs to the lowest energy domain. A total of 99 telescopes are planned to be built in the south and 19 in the north site\cite{sciencewithcta}. This huge step in comparison to current instruments comes with a big challenge in designing, building and maintaining the telescopes.

Here we propose a solution for monitoring the MST structure based on the Operational Modal Analysis (OMA) technique. The technique lies on vibration measurements, the estimation of the telescope eigenfrequencies and eigenmodes and their monitoring throughout the time. The method was applied to the MST prototype structure, which was used for a many-fold of system tests between 2014 to 2019 in Berlin, Germany. The hardware used in the data acquisition and the telescope are described in Section \ref{sec:hardware}. The method is introduced in Section \ref{sec:oma} with its pipeline applied to our case. The results of the monitor are shown in Section \ref{sec:monitoring} and the proof of concept in Section \ref{sec:ropes}. The conclusion and prospects are in Section \ref{sec:conclusion}.

\section{hardware}
\label{sec:hardware}
\begin{figure}
\centering
\includegraphics[scale=1]{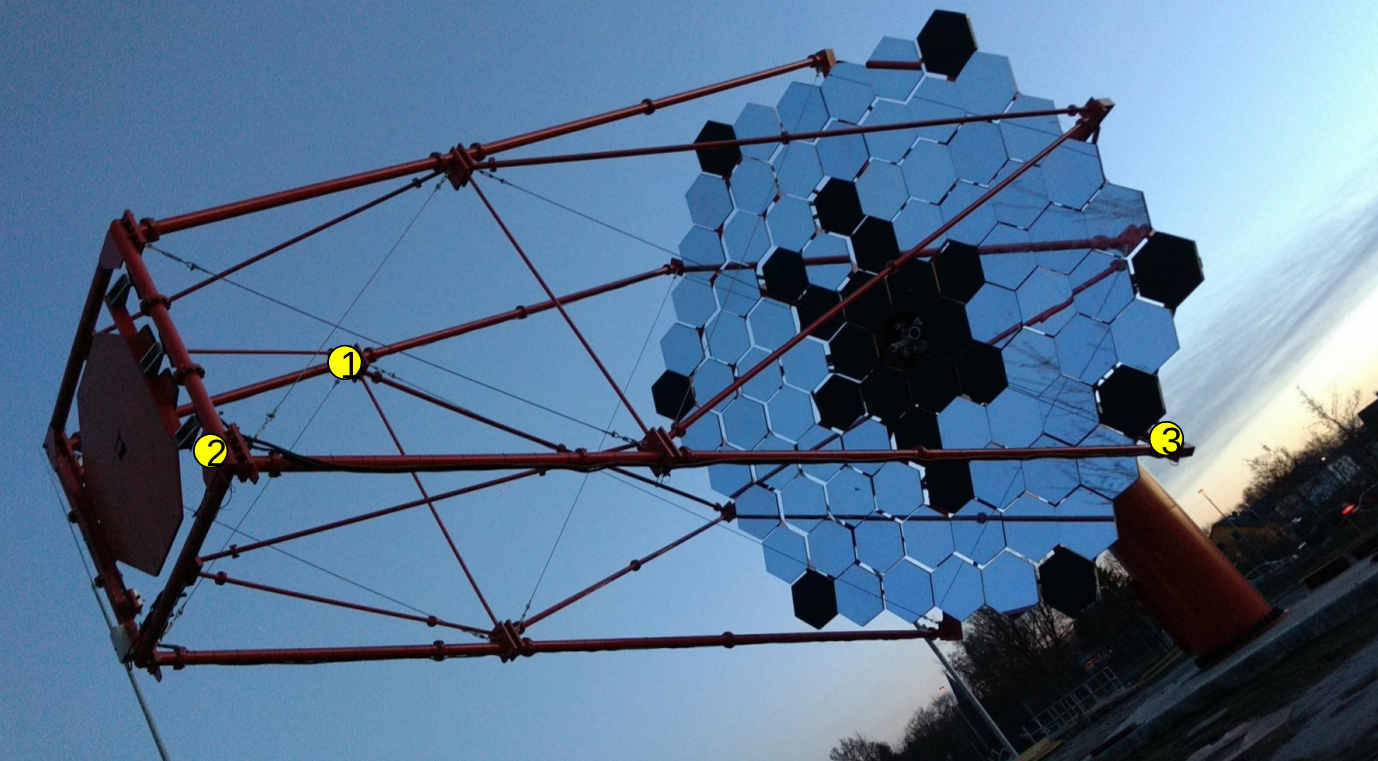}
\caption{Defined positions for the structure accelerometers based on the OMA.}
\label{fig:mst}
\end{figure}

The MST is based on a modified Davies-Cotton design with a reflector diameter of 12 m \cite{markusicrc}. The telescope is composed of 4 parts: tower, dish, Camera Support Structure (CSS) and the Cherenkov camera. A prototype was built in Adlershof, Berlin in 2013 for tests and serial production preparation of mainly the optical and mechanical systems. Fig. \ref{fig:mst} shows the prototype, in this snapshot without the Cherenkov camera. A dummy load is mounted to the camera frame to replace the camera weight. Both are supported by the CSS and its tensioning ropes. A weather station was also mounted near the telescope to monitor the environmental conditions.

The condition monitoring system for the telescopes structure are part of a larger monitoring system developed for the MST prototype. Further monitoring concepts were developed for instance the drive monitoring system and for the bending model. The description of those are presented elsewhere \cite{victoricrc}. 

The measurement of pointing-model parameters and motor torques is possible without dedicated hardware. However, additional hardware is necessary for the measurement of structure vibration (for OMA) and the vibration of motor and gear components of the drive system.

Force balance accelerometers are selected for the vibration measurements. This kind of sensor is widely used for earthquake measurements and at monitoring systems in large structures in the civil industry. In the case of the telescope, it is important to use a very sensitive sensor (high dynamic range) due to the exciting force, the wind, which applies a rather weak force on the structure. See the details of the sensor in Table \ref{table:sensor}.

\begin{table}[]
    \begin{center}
        \begin{tabular}{|c|c|}
            \hline
            \textbf{Vendor}	& GeoSIG\cite{geosig} \\
            \textbf{Model}	& AC-73	\\
            \textbf{Technology}	& Force Balance \\
            \textbf{Frequency range}	& DC to 200 Hz \\
            \textbf{Acceleration range}	& 0 +- 2g \\
            \textbf{Output signal}	& 20 Vpp \\
            \textbf{Operating temperature} & -20$^oC$ to 70$^oC$ \\
            \textbf{Enclosure rating}	& IP68 Z \\
            \hline
        \end{tabular}
    \end{center}
    \label{table:sensor}
    \caption{ Detailed information of the accelerometers selected for the MST structure monitoring }
\end{table}

The accelerometers are mounted at different locations along the CSS to maximize the effectiveness of the method, as it is shown in Fig. \ref{fig:mst}. The monitoring cabinet with the data acquisition is located inside the telescope tower.

Data acquisition of the analog accelerometer output signal is performed with a commercial system from Gantner Instruments\cite{gantner}. Gantner Q.bloxx modules are used for the anti-alias filtering and AD-conversion of the analog sensor signals. A Gantner Q.station provides control over the data acquisition and an ethernet connection for the data transfer.

Data-taking for all measurements is controlled via Alma Common Software (ACS) in dedicated monitoring runs. A special ACS component controls the Gantner Q.station. Data transfer is realized via the FTP protocol to a dedicated FTP-server which forwards data to a noSQL database server running mongoDB. The initial configuration of the Q.station and the connected Q.bloxx modules were only possible using the Windows OS. A Linux version of the Gantner configuration environment is already available for a newer version of the module.

\section{OPERATIONAL MODAL ANALYSIS (OMA)}
\label{sec:oma}

The OMA is an analysis method which focus on estimating the modal parameters of large structures, for instance a telescope. The estimation is based on the measurement of accelerometers mounted on pre-selected locations of the structure. Every rigid body has its own modal parameters, which depend  basically on the geometry and stiffness. These modal parameters will change whenever there is a change in the structure. Changes such as settlement, bending, loosening of screws, material fatigues and relaxation of the tension in the CSS ropes should be reflected in the modal parameters. Therefore, by monitoring of the modal parameters one monitors also the condition of the structure \cite{victoricrc}.

On the other hand, a Finite Element Method (FEM) simulations do provide the expected modal parameters for an ideal structure under predefined conditions. While the FEM simulation does not take into account inhomogeneities in the material, welds and issues during production, the OMA  has a better clue on the real state of the structure within the capabilities of the method. Furthermore, the comparison between simulated and estimated modal parameters is important to assure a good understanding of the structure behavior and ultimately the fulfillment of the requirements.

\subsection{The method}
\label{subsec:method}

The prerequisites for the application of the OMA are:
\newline
-	The excitation on the structure shall be broadband and homogeneous. The input force, for instance the wind, must not be on specific frequencies but should rather cover a large range of frequencies. Furthermore, the force should be applied from different directions and on the structure as a whole;
\newline
-	The accelerometers shall be spread throughout the structure in a way to cover the whole structure. They shall not be concentrated in specific regions or in substructures (such a single beam);
\newline
-	The accelerometers shall be sensitive enough to detect the vibration caused by the input force.

The prerequisites above assure that all the modes of the structure are excited and detected.

The method assumes that the input force is a white noise applied from every direction and in every position of the structure, which is in reality not quite true. The wind speed and direction vary in every timescale.  This variation is though not a big problem since it only influences the intensity and not the position of the modal frequencies in the frequency domain and the detection of higher and less important modal frequencies. To mitigate this effect two approaches were followed: data is taken during a longer period (about one hour) and data quality criteria based on the wind direction and speed were defined. Details follow in the subsequent sections.

The larger the number of sensors available the more modes (and more complex ones) are detected. Consequently, changes in single complex modes, for example, could be interpreted as a well localized damage in the structure. Redundant information is automatically discarded by the analysis method. On the other hand, there is a trade off with the price of the system, since sensors with enough sensitivity to detect vibration caused by wind can be quite expensive. Therefore, the second prerequisite in the list above is actually limited by the price of the sensors. It was found that a three tri-axis sensors configuration (9 channels, ie. 9 degrees of freedom) is enough to derive the main modal parameters of the structure at a reasonable price.

To monitor the modal parameters through time, one shall take data in the exact same structure, otherwise the results will be biased. In the case of the telescope, the exact same structure means that the same elevation and azimuth angle should be used during data taking. For other angle configurations, the structure has a different geometry and, therefore, different modal parameters. An azimuth angle equal to 0° and elevation angle equal to 0° were defined as the standard configuration for data taking for the MST prototype. Environmental conditions such as the seasonal temperature and humidity variation should also influence the estimation of the modal parameters, though in a very small scale compared to the expected change due to structural changes.

\subsection{Rules of thumb}
\label{subsec:rules}
To maximize the quality of the method results the following rules of thumb are suggested, according to \cite{fdd} and references in \cite{svibs}:
\newline
-	The sampling frequency should be larger than 2.5 times the maximum frequency of interest;
\newline
-	The total time of acquisition (s) should be at least 1000 divided by the minimum frequency of interest.

According to MST FEM simulations, the first frequencies of the structure are expected in the range from 1 to 10 Hz. Therefore, the minimum acquisition time should be 1000 seconds and the sampling frequency should be at least 25 Hz \cite{fdd}. From tests on the prototype, it was concluded that a larger sampling frequency (oversampling) helps by better resolving the peaks and a longer acquisition helps by lowering the noise level in the frequency domain. Therefore, the sampling frequency was defined to 100 Hz (which gives a Nyquist frequency of 50 Hz) and the acquisition time to 3750 s.

\subsection{Pipeline}
\label{subsec:pipeline}
The OMA applied for the condition monitoring of the MST is based on the Frequency Domain Decomposition (FDD) technique \cite{fdd}.  The analyses code, developed in Python, is described here with example results for better understanding.

\begin{figure}
\centering
\includegraphics[scale=0.55]{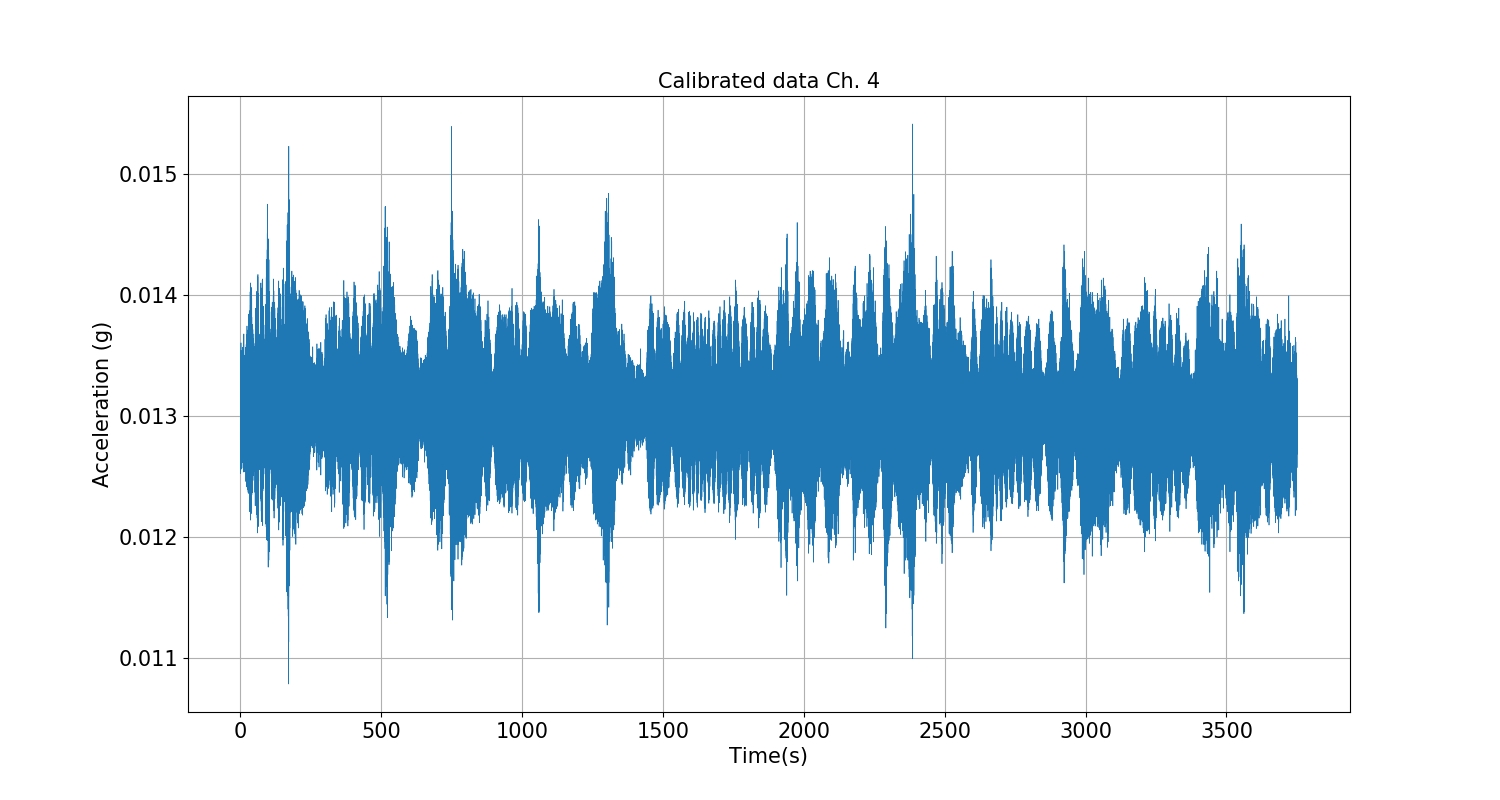}
\caption{Calibrated data for one sensor channel in the camera frame. Data taken on August 25th 2019. The vibration is caused by the combination of all kind of excitation with wind being the main component. The baseline of the signal depends on the orientation of the sensor in regard to the gravity force. The systematic error of the sensor is of the order of 0.1 µm/s².}
\label{fig:input}
\end{figure}

\begin{figure}
    \centering
    \includegraphics[scale=0.35]{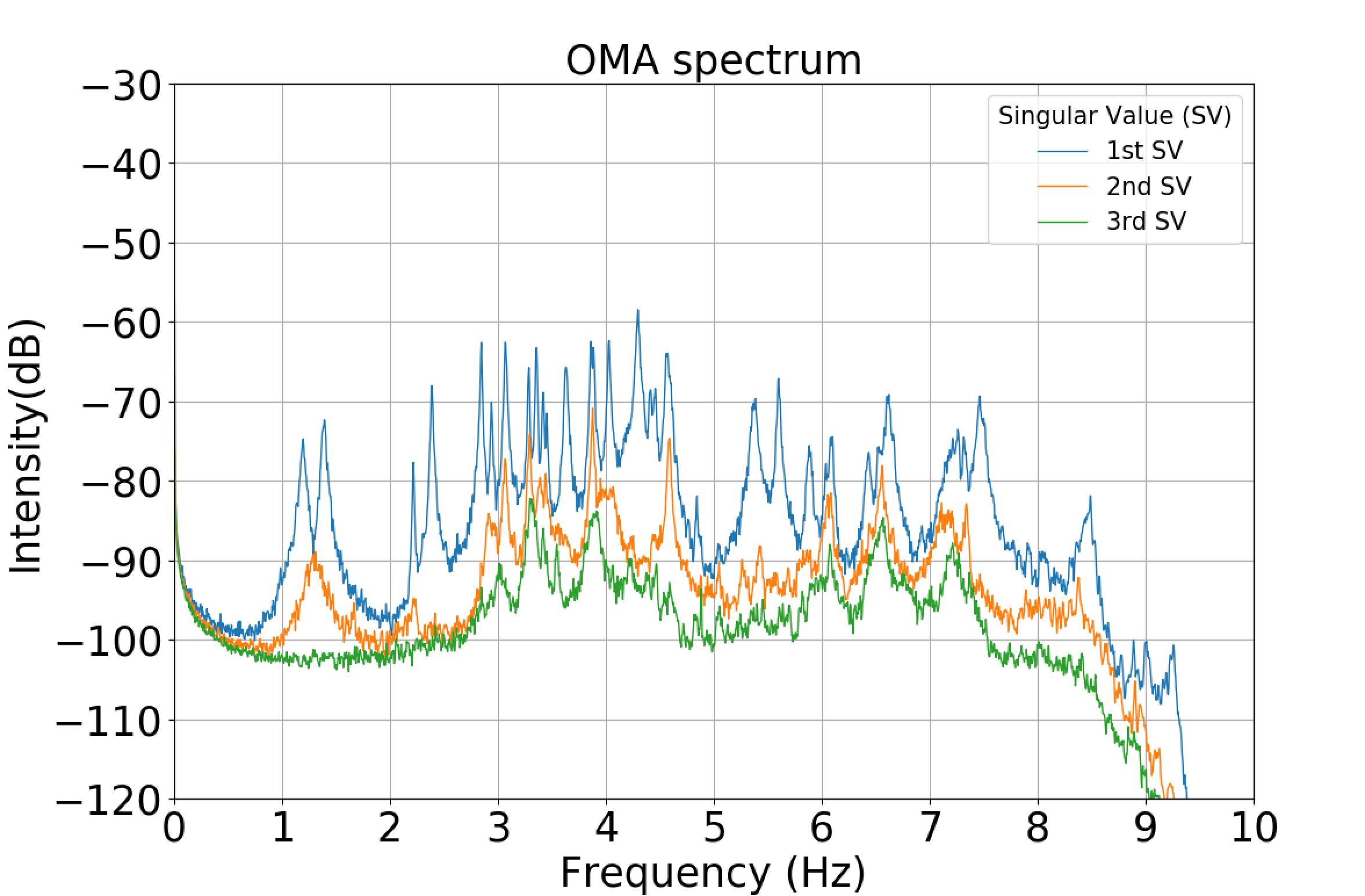}
    \caption{SVD result for data taken on August 28th, 2019. The blue , orange and green are respectively the first, second and third singular values.}
    \label{fig:fdd}
\end{figure}

The pipeline in the analysis code is described below.
\newline
-	Data conversion: using the data-sheet from the sensors the data are converted from mA to g (and m/s²), where g is the gravity constant on Earth. Fig. \ref{fig:input} shows an example of data taken on August 25th. Ch. 4 represents the Y-axis (horizontal movement) for the sensor in the Cherenkov camera frame. \\
\newline
\newline
-	Data decimation: this is applied to reduce the Nyquist frequency from 50 Hz to the maximum frequency of interest (10 Hz). The advantage of oversampling and applying a decimation afterwards is that it resolves better the peaks in the frequency domain and reduces the noise level;\\
\newline
-	Cross Spectral Density (CSD) estimation: it estimates the CSD using the Welch’s method ie. the time series of each channel is converted to the frequency domain by a Discrete Fourier Transform (DFFT) and then individually multiplied by all the other DFFT channels. The result can be gathered together to form a 9x9 matrix for each frequency bin;\\
\newline
-	Singular Value Decomposition (SVD): the 9x9 matrix from the last step is decomposed in three matrices according to the SVD method \cite{fdd}:\\

\begin{equation}
\label{eq:csd}
CSD = U\cdot S\cdot V,
\end{equation}

\noindent where CSD is the initial 9x9 matrix for a frequency bin, U is a 9x9 matrix, which contains the information about the mode shapes; S is a 9x9 diagonal matrix, which contains the singular values of the CSD in decreasing order; and V is usually the transposed U. If the analyzed frequency is a modal frequency, the first value in the diagonal matrix S will be much larger than the subsequent singular values. In this case, the frequency is a modal frequency and the corresponding modal shape is the first column of the U matrix. Fig. \ref{fig:fdd} shows the result of the SVD for the data from the 3 sensors taken on August 25th 2019: the blue, orange and green curves are the first, second and third singular value, respectively. The peaks in the first singular value curve represent the potential modal frequencies. The peaks in the second singular value curve show whenever there is a change in the dominant mode. A peak in the further curves would indicate the degree of degeneracy of that frequency.

\begin{figure}
    \centering
    \includegraphics[scale=0.35]{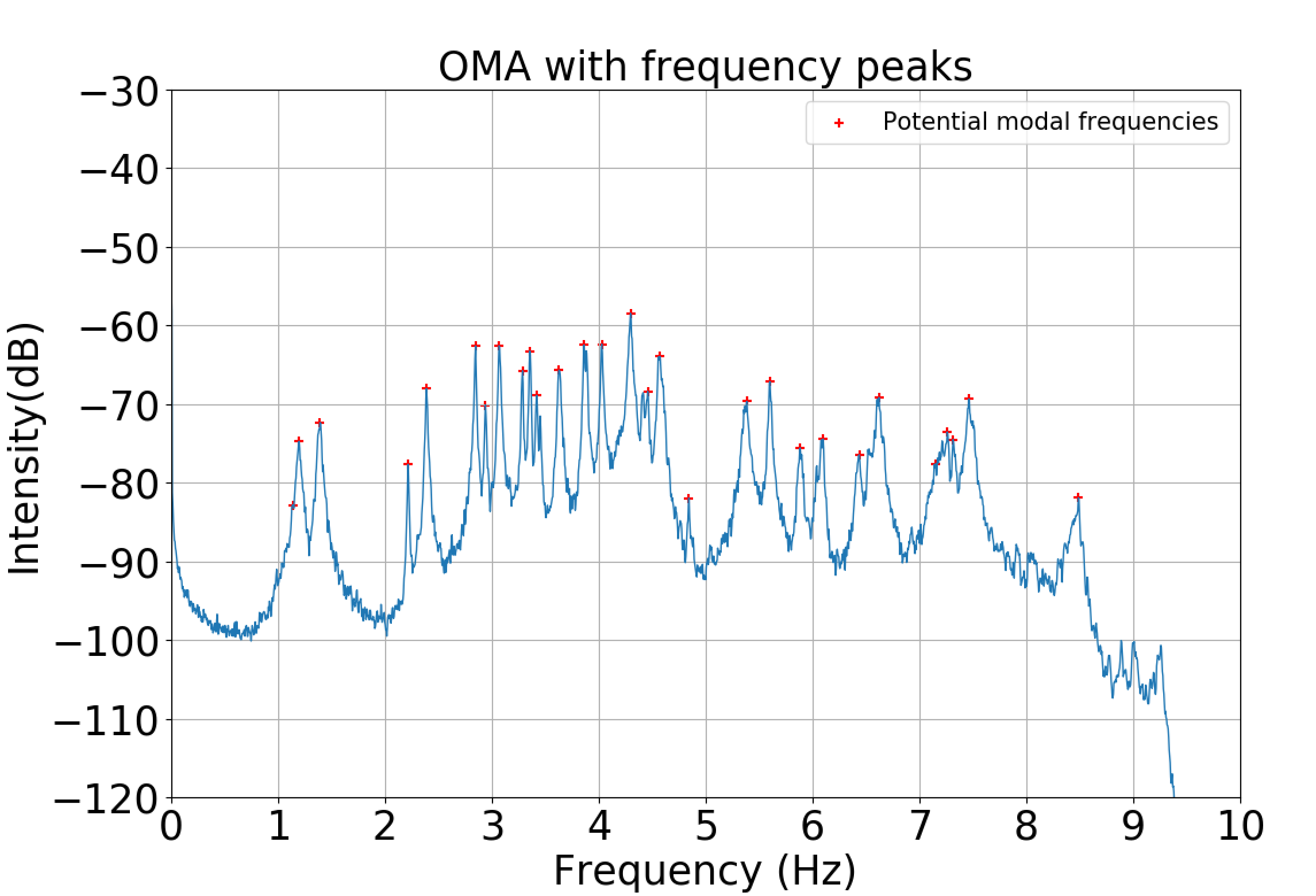}
    \caption{Peak selection in the first singular value curve.}
    \label{fig:fddpeak}
\end{figure}

-	Peak selection: a simple algorithm is applied to the array of first singular values to extract the frequency bins, in which the peaks are located. These peaks are the potential modal frequencies. Fig. \ref{fig:fddpeak} shows the selection of the peaks. It is not a problem when the algorithm takes a fake peak, for instance the first peak in the figure, because these kind of peaks will be excluded later on in the assurance criteria;\\

\begin{figure}
    \centering
    \includegraphics[scale=0.45]{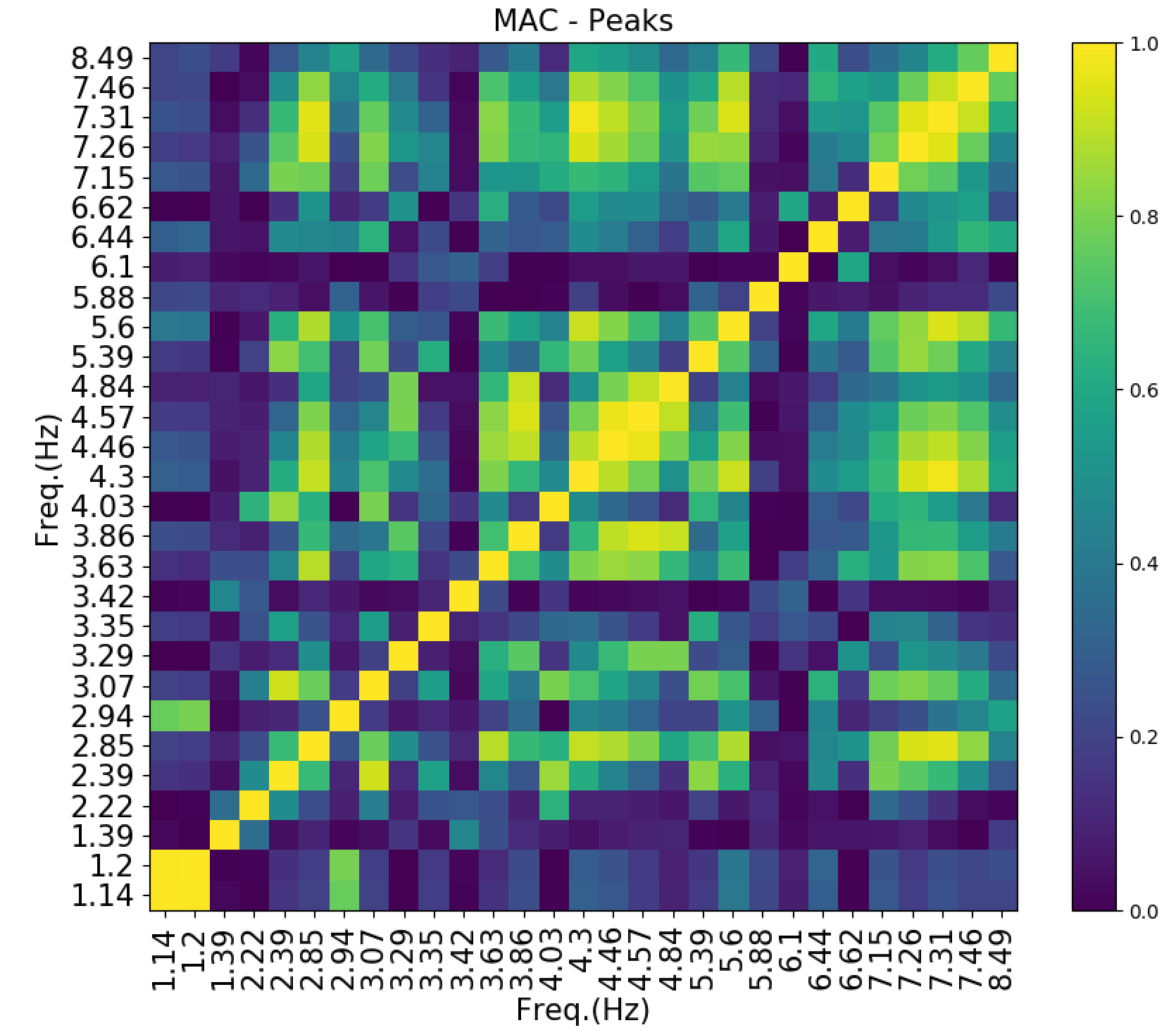}
    \caption{MAC matrix for the selected peaks..}
    \label{fig:mac}
\end{figure}

-	Modal Assurance Criteria (MAC): The MAC value is a value between 0 and 1 to measure how much linearly dependent two vectors are (0 being linearly independent). It is defined by the inner product between one mode shape (9-dimensional vector) and the other, divided by both norms. The MAC value is calculated for every pair of potential modal frequencies found in the last step. The result is a MAC matrix showing the linearly dependency among all the encountered peaks. Fig. \ref{fig:mac} shows the MAC matrix for all the selected peaks in the example case (August 25th, 2019). As pointed out before, it was identified that the first two peaks are strongly correlated. Many other correlations between the peaks were also identified and are treated in the next step;\\

\begin{figure}
    \centering
    \includegraphics[scale=0.45]{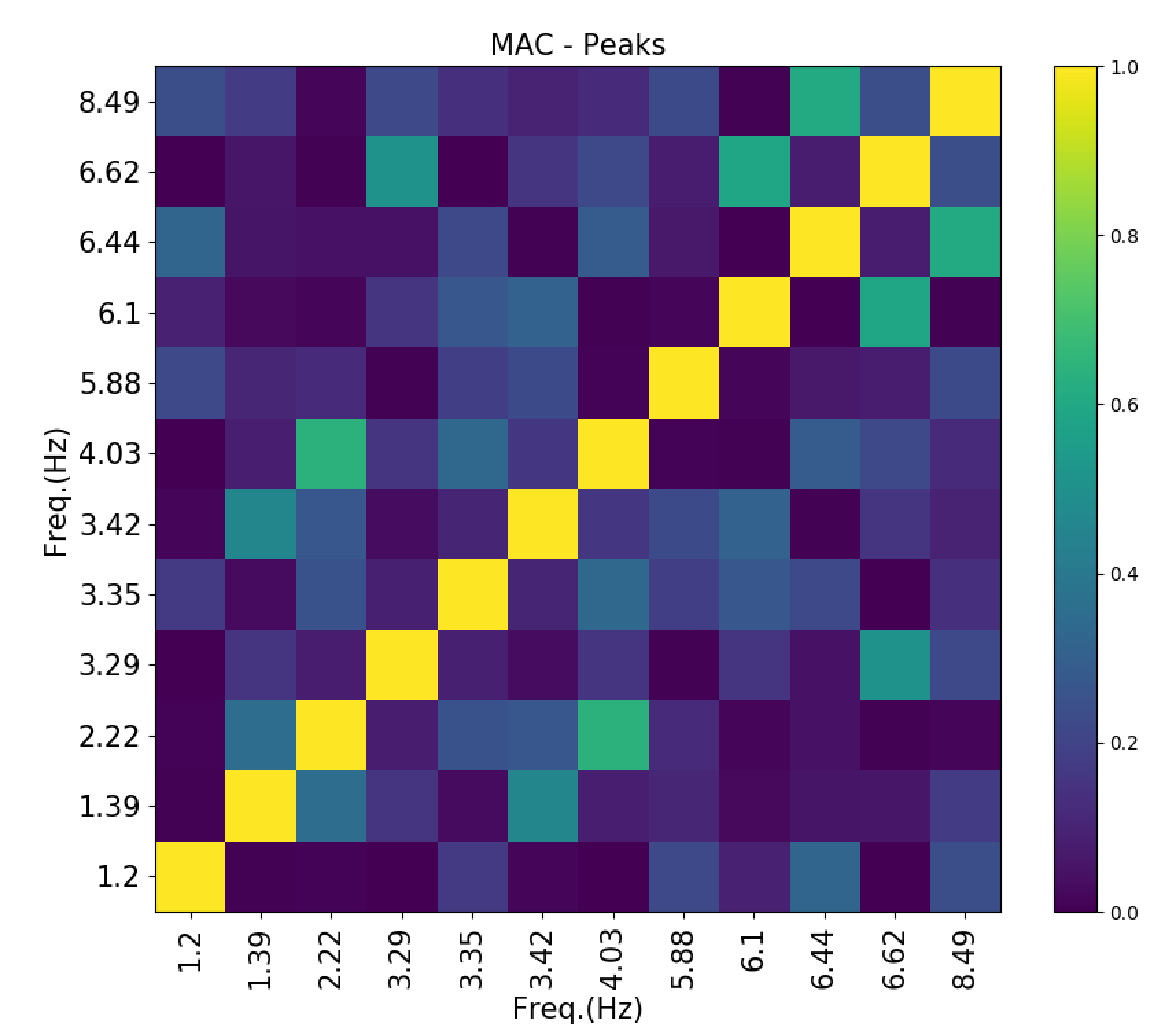}
    \caption{Cleaner MAC matrix after excluding correlated modes.}
    \label{fig:mac2}
\end{figure}

-	MAC selection: the MAC value of 0.8 is defined as the maximum accepted value for two mode shapes to be considered independent from one another. If the MAC value is larger than 0.8, the two mode shapes are considered to be linearly dependent and the one with larger sum of all MAC values in respect to the other vectors is discarded. Finally, the result is a new and cleaner MAC matrix together with the list of the structural modal frequencies. Fig \ref{fig:mac2} shows the new matrix.\\

\begin{figure}
    \centering
    \includegraphics[scale=0.4]{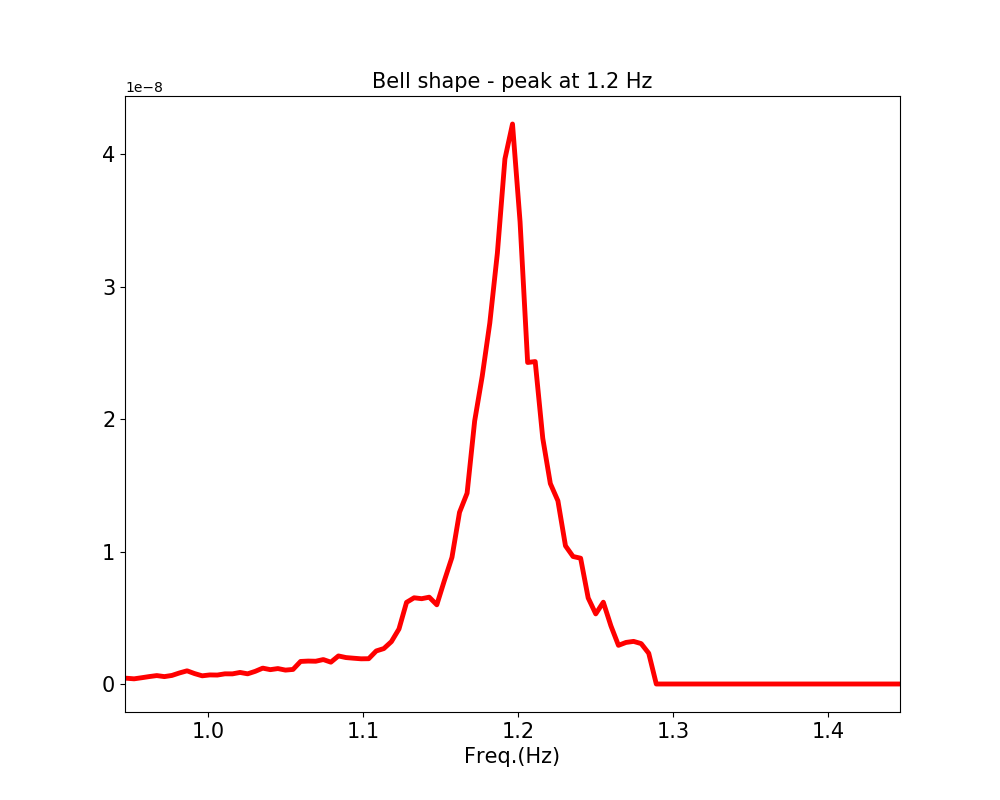}
    \caption{Bell shape curve around the first modal frequency.}
    \label{fig:bell}
\end{figure}

-	Bell shape curve: for each modal frequency, a second MAC value is calculated between the central frequency bin where the peak is located and the nearby ones (by using the first column vector of U in Eq. \ref{eq:csd}). Whenever the MAC value is larger than 0.95 the frequency bins are considered to be part of the same mode  and the intensity of that bin is kept the same as before (the first singular value for that frequency bin). For all the other frequency bins the intensity is set to zero. The result is a bell shape curve shown in Fig. \ref{fig:bell} for the first modal frequency found before. The intensity unit is not important for the following analysis. \\

\begin{figure}
    \centering
    \includegraphics[scale=0.4]{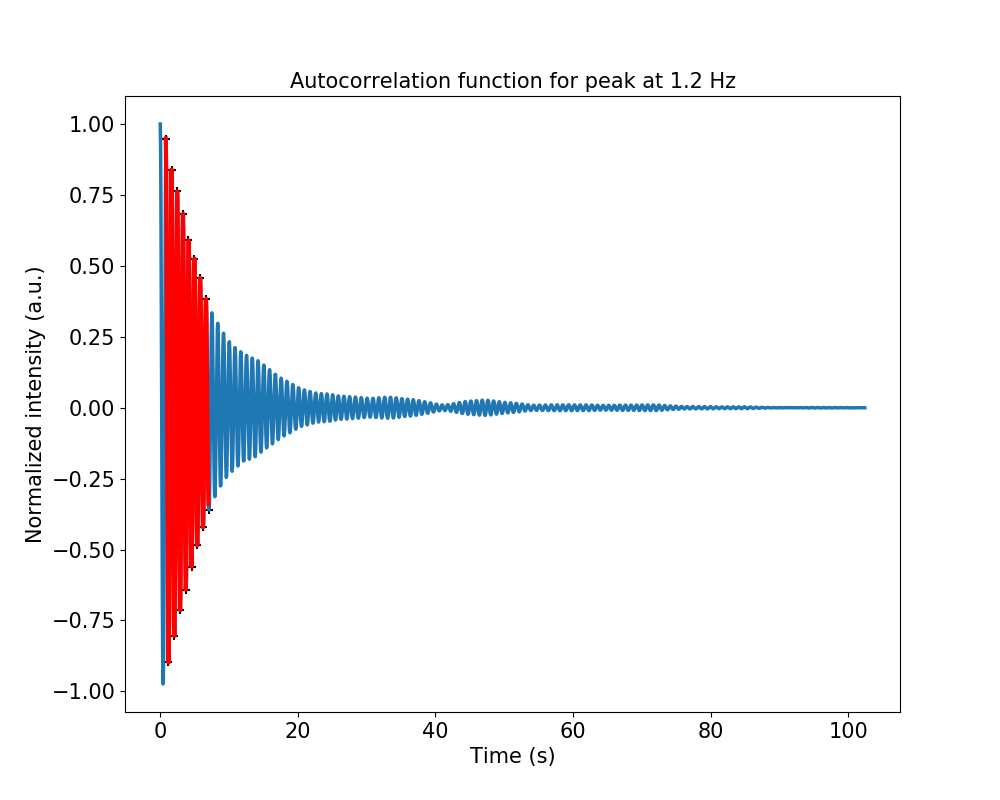}
    \caption{Auto correlation function for the first modal frequency. The blue curve is the decay curve, the red curve is the region of interest (between 0.3 and 0.7 a.u.) and the black crosses indicate the extremes position.}
    \label{fig:autocorr}
\end{figure}

-	Auto correlation function: the auto correlation function for each modal frequency is defined by the Inverse Fourier Transform (IFFT) of the bell shape curve. It is then normalized to maximum intensity 1 and average 0. The function represents the decay curve of the vibration at that modal frequency. Fig. \ref{fig:autocorr} shows the auto correlation function for the first modal frequency;\\

\begin{figure}
    \centering
    \includegraphics[scale=2]{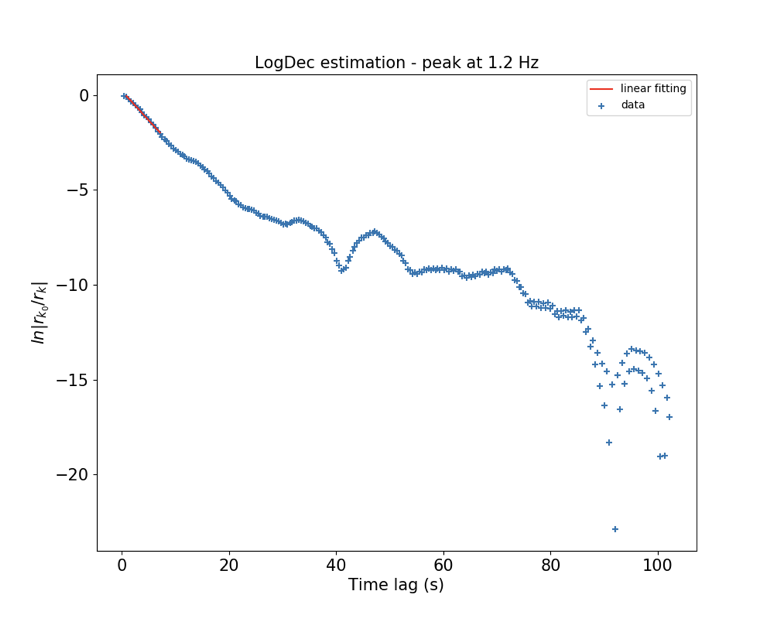}
    \caption{The logarithm decimation function for the first modal frequency. Blue dots show the estimated function and the red line the linear fit within the region of interest.}
    \label{fig:logdec}
\end{figure}

-	Logarithm Decimation: the logarithm decimation function is defined as twice the logarithm of the relation between the intensity of the first extreme (peaks and valleys) and the n-extreme in the auto correlation curve. The logdec factor $\xi$ is defined as the slope of the intensity of this function as function of the number of the peak. The range to be considered in the linear fitting is the region where the norm of the auto correlation function is between 0.3 and 0.7. This region of interest avoids taking into consideration regions close to the start and end of the excitation, where non linear effects could lead to more uncertainties in the estimation. Fig. \ref{fig:logdec} shows the function also for the first modal frequency.\\

Figs. \ref{fig:logdec} and \ref{fig:autocorr} show that the behavior of the oscillation is only linear within the first beating envelopes.

-	Damping rate estimation ($\partial$): Finally, based on the logdec factor $\xi$, the damping rate for each modal frequency can be estimated through the equation \ref{eq:logdec}\\

\begin{equation}
\partial = \dfrac{1}{\sqrt{1+(\dfrac{2\pi}{\xi})^2}}.
\label{eq:logdec}
\end{equation}

The damping rate is usually estimated with large uncertainties, since there are some assumptions before reaching the final estimated value. These uncertainties can be identified as the standard deviation of the damping rate of each modal frequency throughout the time if no change is expected in the structure (see Section \ref{sec:monitoring}).

Table \ref{tab:results} summarizes the results for the August 25th data. A crosscheck analysis with an independent commercial software (Artemis Modal from Svibs \cite{svibs}) delivered similar results.

The estimated mode shapes can be better visualized through the commercial software Artemis Modal and are also presented in Table \ref{table:freq}. It is not simple to interpret the mode shape as a simple translation/rotation, because the real structure is more complex than its CAD design. Effects such as small asymmetries, different screw torques, sagging and other possible deformations makes this interpretation difficult, hence the third column of the table might be inaccurate.

\begin{table}[]
    \label{table:freq}
    \begin{center}
        \begin{tabular}
        {|c|c|c|}
        \hline
        \textbf{Modal Freq.(Hz)} & \textbf{Damping(\%))} & \textbf{Mode shape }\\
        \hline
        1.20	& 2.04	& Translation (Y-axis) \\
        1.40	& 1.40	& Translation  (X-axis )\\
        2.22	& 0.36	& Rotation around Z \\
        3.29	& 0.19	& Rotation around X\\
        3.35	& 0.22	& Translation (Z-axis)\\
        3.42	& 0.22	& Translation  (mostly X-axis, complexer mode)\\
        4.02	& 0.25	& Rotation around Z\\
        5.90	& 0.27	& Rotation around Y\\
        6.07	& 0.17	& Translation  (X-axis )\\
        6.40	& 0.29	& Translation (Z-axis)\\
        6.59	& 0.3	& Rotation  around X\\
        8.46	& 0.24	& Rotation around Y\\
        \hline
        \end{tabular}
    \end{center}
    \caption{Summary of the OMA results for data taken on August 25th. X-axis is the gravity direction, Y-axis the horizontal direction and Z-axis the telescope axis}
    \label{tab:results}
\end{table}

The advantage of the analysis code presented here in comparison to the commercial software is the automation and customisation, which are important features in the tracking of the modal parameters for a large number of MST telescopes over long time scales.

\section{LONG TERM MONITORING}
\label{sec:monitoring}
The OMA results shown in Section \ref{subsec:pipeline} are for data taken on August 25th, 2019.  The main purpose of the analysis is to develop a monitoring system, which tracks the modal parameters throughout the time and detects changes in the structure. In this section the results of the automated analysis code for the tracking are presented.

Data was taken automatically every day at 6:00 a.m. at the MST prototype in Adlershof, Berlin. The data acquisition is defined according to Section \ref{sec:hardware}: sampling ratio of 100 Hz, acquisition time of 3750 s, telescope at 0° elevation and 0° azimuth. The three tri-axis sensors and the whole system were working properly from August 16th, 2019 until December 26, 2019 except between August 26th and August 28th, when there was an update in the system. The analysis also ran automatically every day. 

To improve the reliability of the results, it is important to apply a selection criteria to the data to cut off all the days in which the excitation force was either too weak or too strong. When the wind is weak, less modes are excited and when it is strong new and more complex modes are excited. When the wind comes mostly from one direction, there is the risk that not all the telescope modes will be excited, therefore a criteria in the wind direction variance was also applied. Based on the data acquired by the weather station, the quality criteria was defined as shown in Table \ref{table:criteria}. The criteria was applied for the time window of the data acquisition (6h - 7h) on each day in which data was acquired.

\begin{table}[]
    \begin{center}
        \begin{tabular}
        {|c|c|c|}
        \\hline
        \textbf{Minimum Mean wind speed($m/s$)} & \textbf{Maximum Mean wind speed($m/s$)} & \textbf{Wind direction variance($^o$)} \\
        \\hline
        1.1	& 2.5	& 120 \\
        \\\hline
        \end{tabular}
        \\hline
    \end{center}
    \caption{Quality criteria}
    \label{table:criteria}
\end{table}

\begin{figure}
    \centering
    \begin{subfigure}
        \centering
        \includegraphics[scale=1]{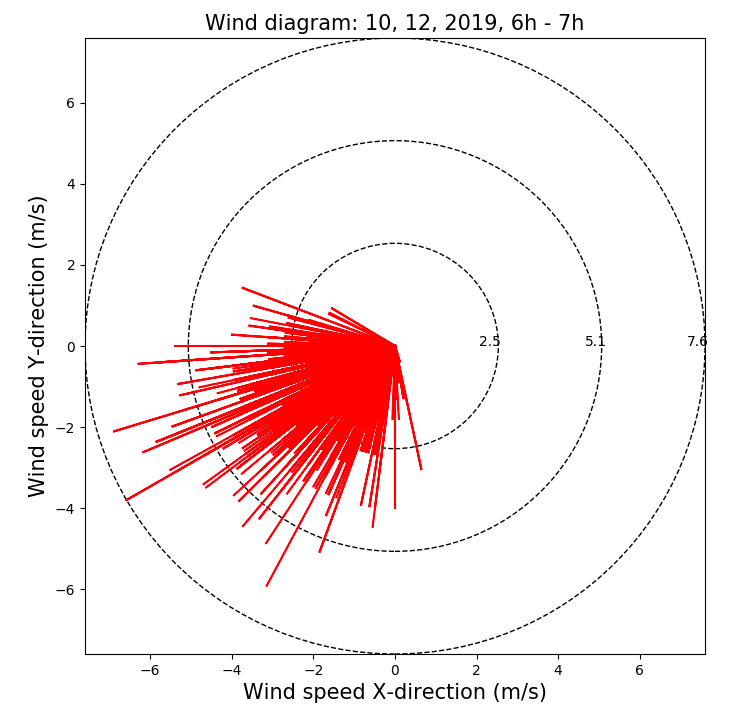}\\
        \label{fig:wind1}
    \end{subfigure}
    \hspace{1pt}
    \begin{subfigure}
        \centering
        \includegraphics[scale=1]{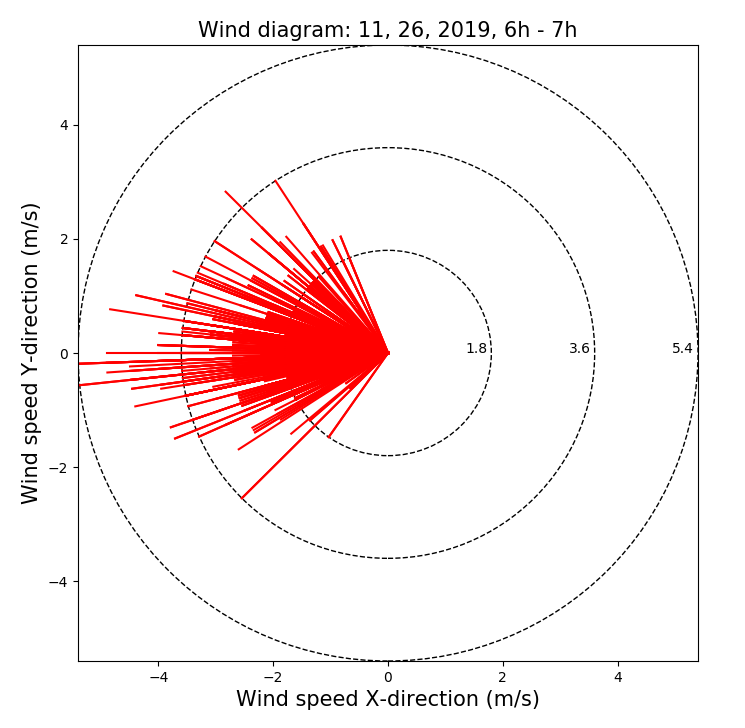}\\
        \label{fig:wind2}
    \end{subfigure}
    \hspace{1pt}
    \begin{subfigure}
        \centering
        \includegraphics[scale=1]{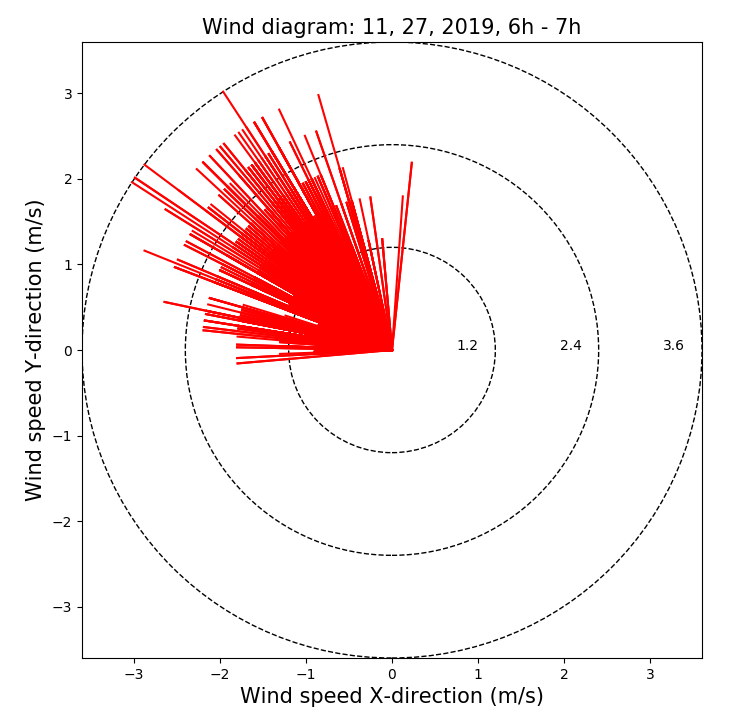}\\
        \label{fig:wind3}
    \end{subfigure}
    \hspace{1pt}
    \hspace{1pt}
    \caption{Wind diagram for three different days: 2019 October 12th (upper panel),2019 November 26th (middle panel), 2019 November 26th (lower panel).The wind strength is given by the length of the red lines which originate at the center where the telescopes are located. The direction is also given in 360$^o$}
    \label{fig:wind}
\end{figure}

Fig \ref{fig:wind} shows the wind information for 3 different days in which the wind passed the quality criteria (see text). From upper to lower panel: a day with strong wind, a rather normal day and a weak wind day.

\begin{figure}
    \centering
    \includegraphics[scale=1.8]{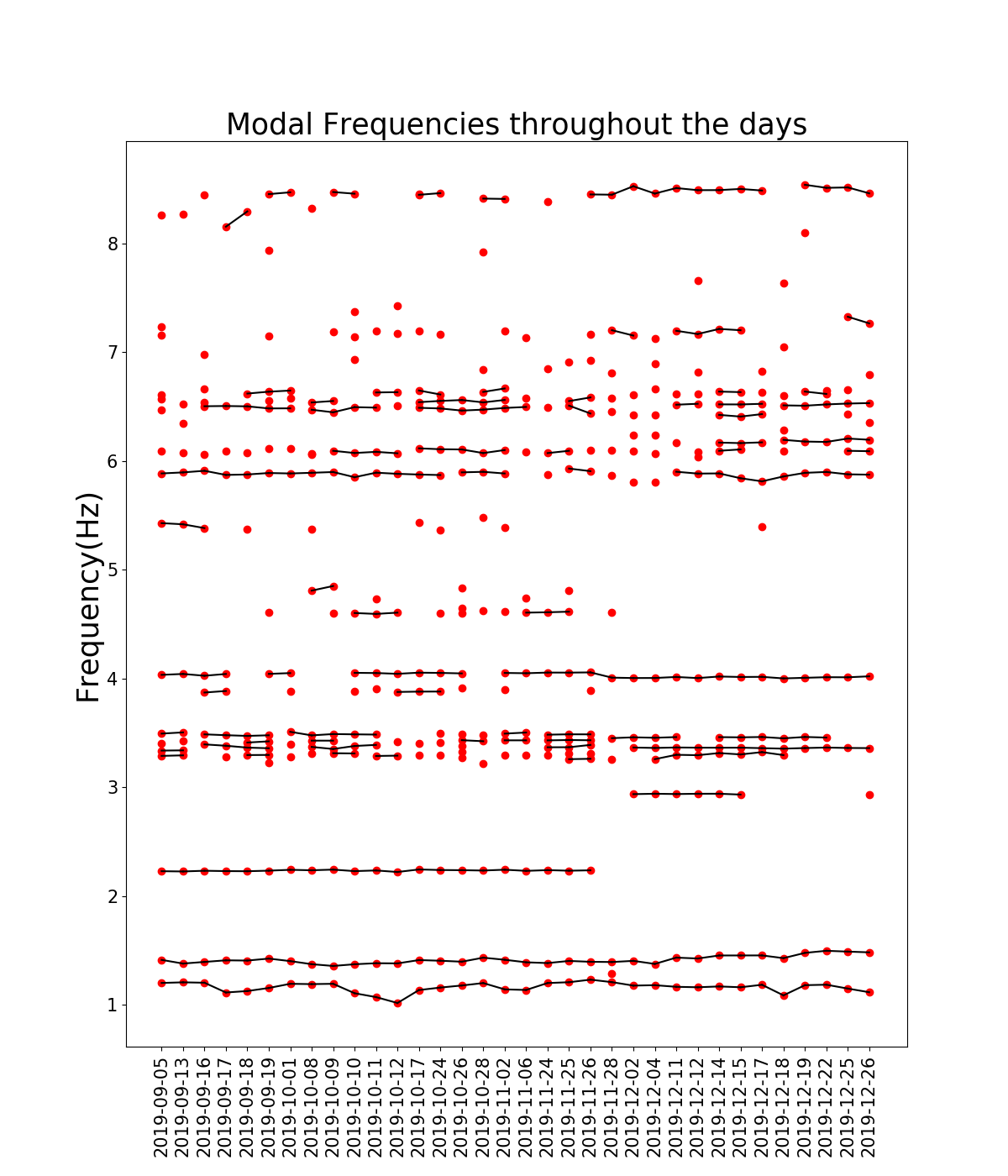}
    \caption{Tracking modal frequencies throughout the days. The red dots are the modal frequencies in each day and a black line is drawn whenever there is a correlation between consecutive day modes.}
    \label{fig:monitor}
\end{figure}

The modal frequencies are shown in Fig. \ref{fig:monitor}. The black lines connect together correlated frequencies for every consecutive day, i.e. a MAC value is calculated between every frequency in one day and every frequency in the consecutive day. A correlation happens whenever the MAC value is larger than 0.95. Some of the dots are not connected, despite being in the same position at the day before, because there was not enough information available for the MAC value to be above the threshold. This can happen for two reasons: either there was a lack of excitation or the mode is of a higher order and is difficult to be excited. The more channels available (the more sensors) the more accurate the MAC value is and the more linearly independent (higher) modes are tracked.

\begin{figure}
    \centering
    \includegraphics[scale=1.8]{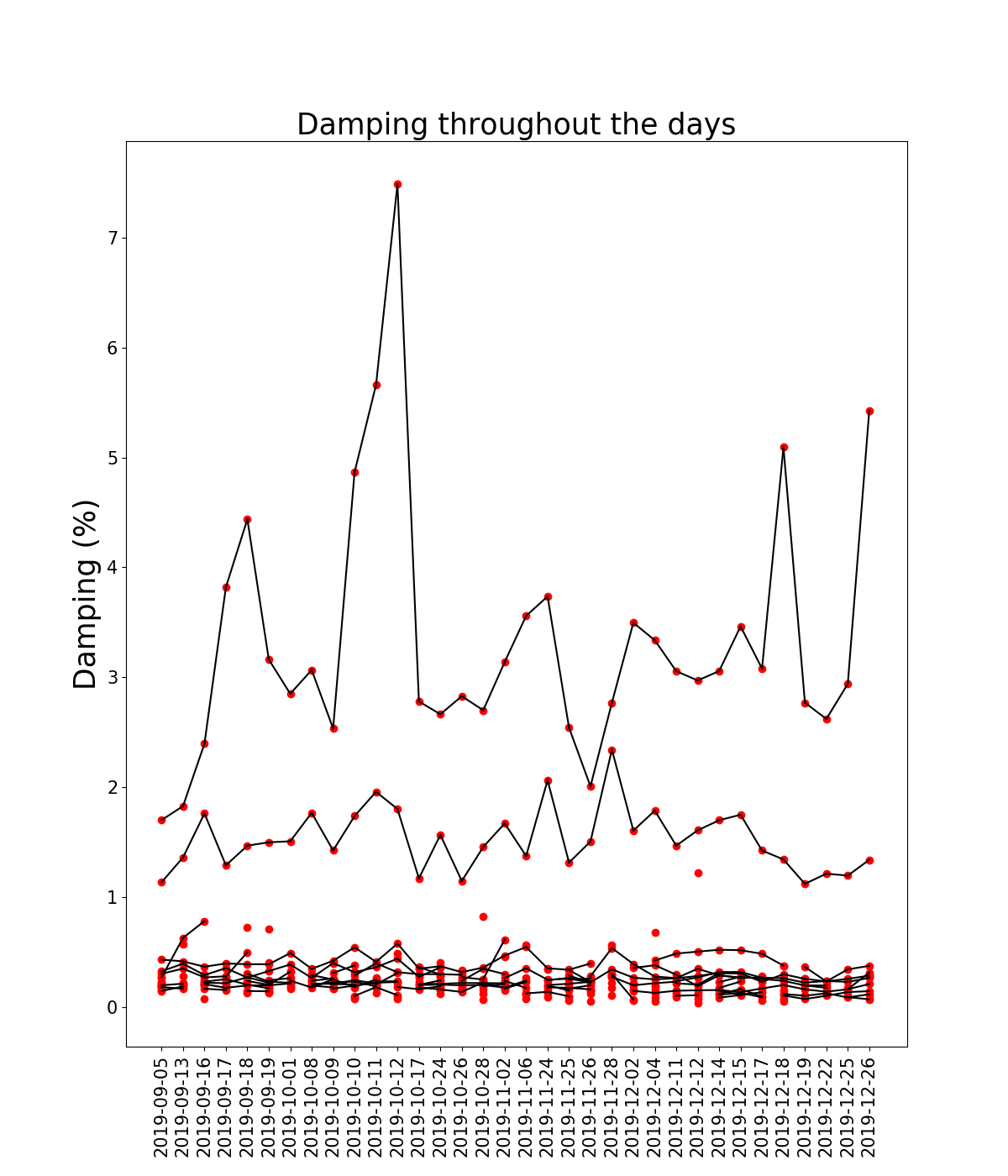}
    \caption{Tracking damping rate throughout the days. The red dots are the damping rates in each day and a black line is drawn whenever there is a correlation between consecutive days modes.}
    \label{fig:damping}
\end{figure}

The damping rate for each modal frequency is also tracked and plotted in Fig. \ref{fig:damping}. As it was mentioned before, the uncertainties are higher than for the Eigenfrequency estimation; therefore, there is a relatively larger standard deviation for correlated modes in comparison to the frequency tracking.

In Fig. \ref{fig:monitor} and \ref{fig:damping} one can identify a special day on October 12th, when the first modal frequency is smaller than the days before and the damping is higher. On the very next days, the frequency and damping of the first mode came back to their usual values, indicating that this was not a change in the structure itself. The reason for this change is in the excitation force. The wind on October 12th between 6h and 7h was in average within the quality limits although there must have been a spike in the wind speed since from Fig \ref{fig:wind} upper panel the maximum value reaches up to 7.6 m/s, while in other days this is not usual. Therefore, there is space for improvement in the quality criteria definition. Furthermore, information about rain rate, temperature, humidity and pressure could have an influence on the results but was not used in the quality criteria definition since these are not the primary sources of excitation.

The last step of the monitoring analysis is to define variables which can suggest whether there was a change in the structure or not. These variables can be defined in many different ways either in the frequency  or in the time domain\cite{damage}. We defined three different variables to be monitored: 1) square frequency change (Eq. \ref{eq:chifreq}): the sum of squared differences between the modal frequencies in subsequent days, 2) square damping change (Eq. \ref{eq:chidamp}): same as 1 for the damping and 3) the number of correlation established between subsequent days (black lines in Fig \ref{fig:monitor} and \ref{fig:damping}). 

\begin{equation}
\Delta f_i ^2 = \dfrac{\Sigma^{N}_j(f_{i+1,j} - f_{i,j})}{N}
\label{eq:chifreq}
\end{equation}

\begin{equation}
\Delta \partial_i ^2 = \dfrac{\Sigma^{N}_j(\partial_{i+1,j} - \partial_{i,j})}{N}
\label{eq:chidamp}
\end{equation}

\noindent where $i$ presents the day, $j$ the jth correlation between $i$ and $i+1$ and $N$ is the total number of correlation between those days.

The first two frequencies in Fig. \ref{fig:monitor} correspond to the two highest damping in Fig. \ref{fig:damping}. These modes are contaminated with a tiny rotation in horizontal and vertical axis, specially seen under strong wind, when the width of the two peaks increase significantly. To overcome this natural barrier for the monitoring, one could define a more strict range for the allowed wind speed or alternatively, do not take these two peaks into consideration when calculating the monitor parameters. Nevertheless, for the further analysis we stick to the current quality criteria defined in Table \ref{table:criteria} and kept those two peaks.

\begin{figure}
    \centering
    \includegraphics[scale=1.4]{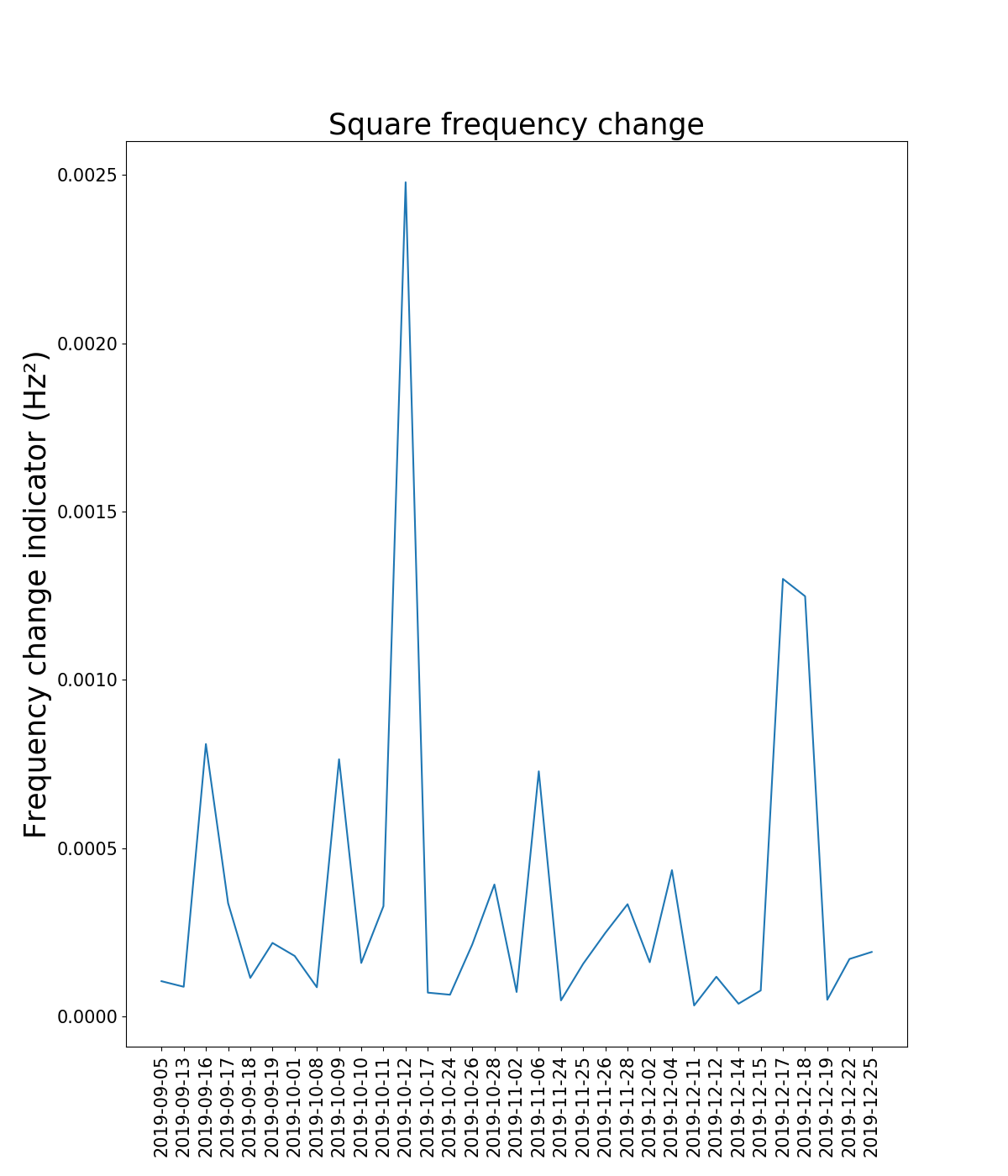}\\
    \hfill
    \caption{Square frequency change indicate how much the modal frequencies of the same mode shape changed between one day and the next. A large peak indicates a change in the structure. For the period analysed the structure is not expected to change.}
    \label{fig:chifreq}
\end{figure}

\begin{figure}
    \centering
    \includegraphics[scale=1.4]{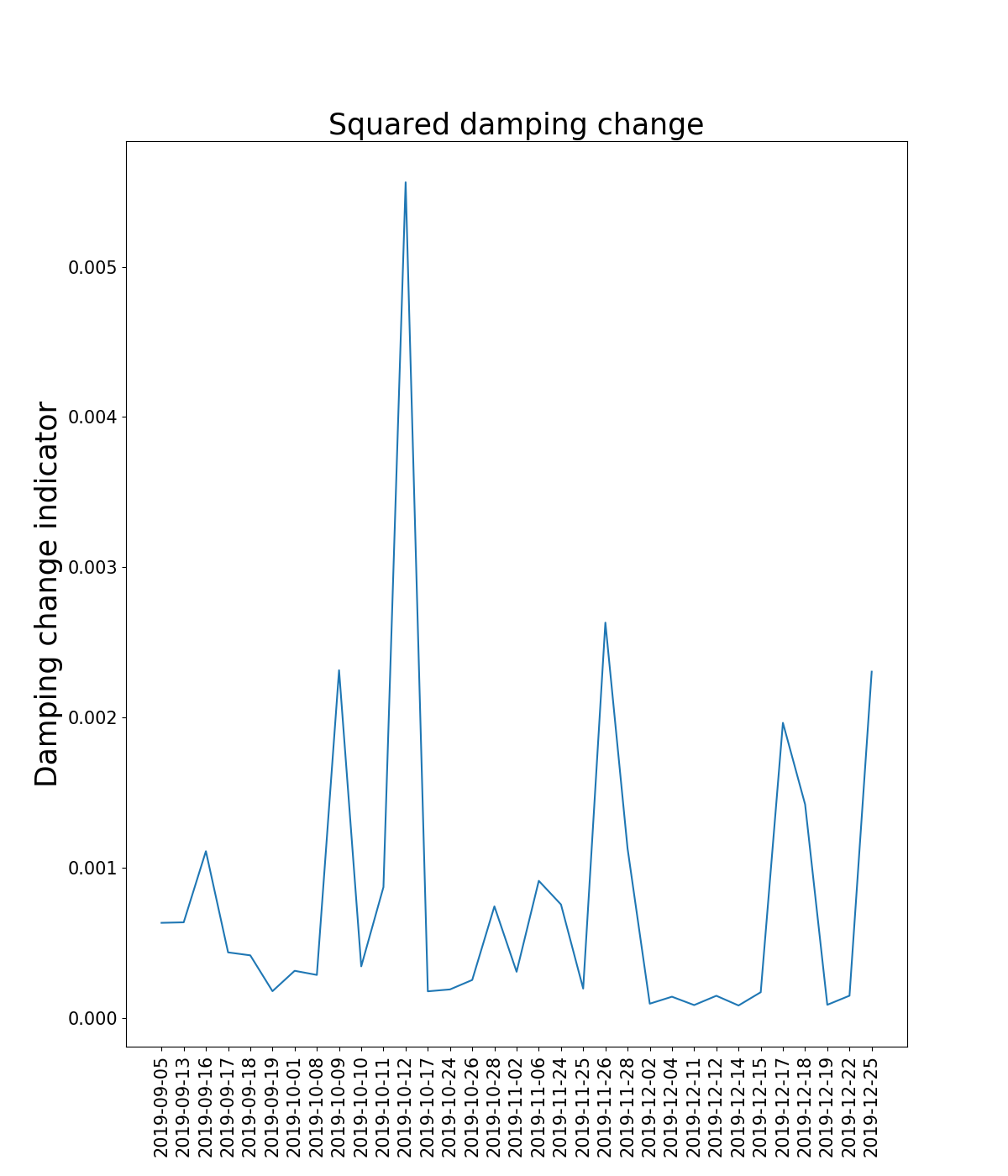}\\
    \hfill
    \caption{Square and Square damping change indicate how much the damping rates of the same mode shape changed between one day and the next. A large peak indicates a change in the structure. For the period analysed the structure is not expected to change.}
    \label{fig:chidamp}
\end{figure}

Fig \ref{fig:chifreq} and Fig \ref{fig:chidamp} show the monitoring of the defined variables $\Delta \partial_i ^2 $ and $\Delta f_i ^2$ for the same period. There is a clear correlation between a change in the modal frequencies and a change in the damping rate, which shows the robustness of the analysis.

In spite of the good correlation between the parameters, at first one cannot be sure when a change in the structure occurs as long as it doesn't happen and its effect is deeply investigated. A threshold must be defined in the parameters of Figs. \ref{fig:chifreq} and  \ref{fig:chidamp} and whenever the calculated parameters exceed this threshold, a change most probably occurred in the structure. In the prototype system developed in Adlershof, an automatic email was sent everyday to the persons in charge with the main plots, tables and a main message telling the reader whether there was a change or not.

\section{PROOF OF CONCEPT}
\label{sec:ropes}
More important than monitoring the modal parameters is to be able to investigate what happens when the structure changes. On November 26th and 27th 2019 a experiment was conducted at the prototype to study the influence of the tension in the CSS ropes at the modal parameters. Special tools were rented from an elevator manufacturer company and used to apply the tension and another one to measure it. Dedicated runs were conducted to extract the modal parameters from the structure during the experiment. To optimize the experiment duration, the dedicated runs had the duration of only 15 minutes instead of the normal 1 hour. The procedure was the following:

\bigskip
\noindent- Before tightening: A dedicated 15 minutes data taking run for extracting the initial modal parameters;\\
- 7.85 kN was applied to each rope when the telescope was pointing towards 90 degree elevation;\\
- A second dedicated 15 minutes data taking run;\\
- Relaxation: the telescope stood still for about an hour;\\
- A dedicated 15 minutes data taking run;\\
- After operation: the telescope was moved in random directions for about 10 minutes;\\
- A dedicated 15 minutes data taking run;\\
- Two cables had their tension intentionally released to 3 kN;\\
- A dedicated run was taken to see the effects of the releasing;\\
- The nominal tension of 10 kN was applied to every rope while the telescope was pointing towards 90 degree elevation;\\
- A dedicated run was taken to see the effects of the tightening;\\
- Relaxation: The telescope stood still until the next morning;\\
- The last dedicated run.
\bigskip

\begin{figure}
    \centering
    \includegraphics[scale=1.2]{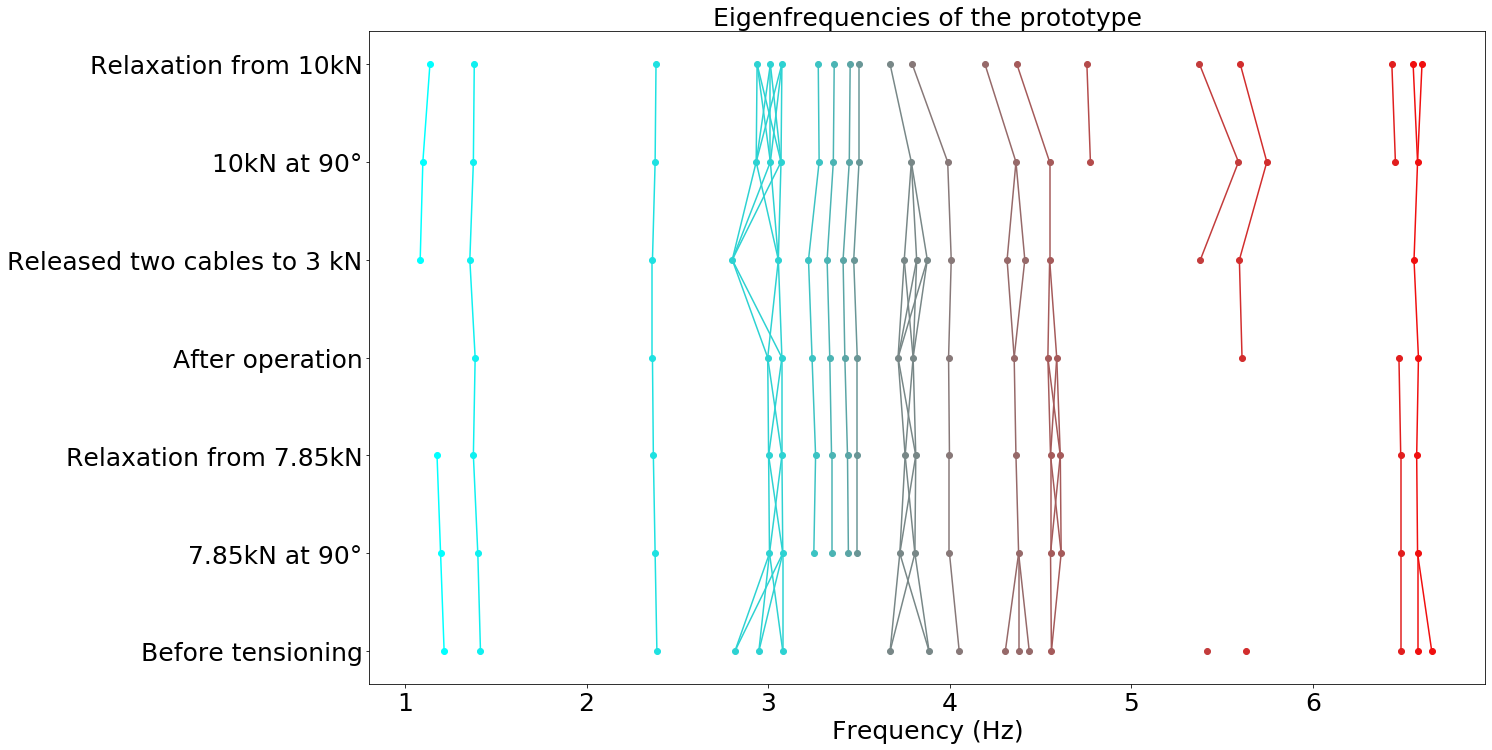}
    \caption{The monitoring of the peak frequencies during the experiment conducted on November 26th and 27th. See the text for more information.}
    \label{fig:ropes}
\end{figure}

Fig. \ref{fig:ropes} shows the results for the experiment. The MAC criteria were relaxed when comparing the different peaks, in comparison to the value used in Section \ref{sec:monitoring}), therefore more peaks could be tracked but with larger uncertainty. A line is drawn whenever there is a correlation between the mode shapes of the subsequent datasets. As explained in the last section, the first two frequencies are not modal frequencies. The peak around 2.4 Hz is present during the whole experiment, although in the long term monitoring it is not anymore identified as an independent frequency after November 27th. The three peaks around 3 Hz merged together when tightening the cables to 7.85 kN, kept the same values during relaxations and split during the release of cables. The frequencies in the region from 3.2 to 3.5 Hz were not strong enough before the first tightening of the cables. During the tightening of the cables to 10 kN these frequencies slightly shifted to higher frequencies. This indicates that the structure became stiffer. Peaks around 4 Hz also merged together when tightening the cables and split when releasing. In general it is also seen that the relaxation from 10 kN was felt by peaks with frequency larger than 3.5 Hz, since they decreased, therefore the structure became less stiff.

\section{CONCLUSION}
\label{sec:conclusion}
The condition monitoring system was developed and tested at the MST prototype in Berlin. Given the large number of telescopes planned for the CTA Observatory, a monitoring of the structures is key to avoid big failures and minimize maintenance. The system is based on the Operational Modal Analysis and on accelerometer measurements. The system developed proved to be functional, since the change in the ropes tension were identified as changes in the modal frequencies. Whenever a tension in the cables is released, some of the frequencies shifted towards lower frequencies and others split into more peaks. The opposite is also true, i.e. the modal frequencies shift towards higher values when the tension is higher and some frequencies tend to come together. By relaxation, some peaks shift towards lower frequencies. With these conclusions, one can monitor the tension in the cables without the need to measure it, which is a big advantage when one deals with dozens of telescopes in the desert. 

Based on the results from the experiment conducted in November, it is also possible to conclude that the first three modal frequencies (in Fig. \ref{fig:ropes} do not belong to the CSS modes but rather from the structure as a whole, since they did not show any shift during the tightening/releasing of the cables. To identify specific modes to track for specific problems is also important for the further development of the system.

More structural changes could have been tested but due to time limitations it was not feasible. It is expected that the system detects overall changes, such as the tension in the ropes, settlement of the ground, ruptures and long term fatigues in the material. It is not expected that the system detects minor, localized issues and dynamic effects such as loosening of screws and level of lubrication of the gears.

Despite the fact that the system is functional, the two defined parameters to monitor the health of the telescope (Eq. \ref{eq:chifreq} and \ref{eq:chidamp}) did not identify any change in the structure before and after the experimented conducted in November 26th. Due to the sum, the effect seen in some frequencies was smoothed out by the other frequencies in which there was no shift. This shows that the definition of the parameter to monitor is important and can be improved for the automatic detection of failures. Although these two parameters were unable to identify the change in the structure, the third one, which is the number of correlation between consecutive days, could give a hint of a change. It is clear from Fig. \ref{fig:monitor} and \ref{fig:damping} that there was a very low number of correlations between November 26th and November 28th. Furthermore, the long sequence of correlations of the 2.15 Hz peak disappeared and a new sequence around 2.9 Hz showed up.

To take advantage of all information the data provides and to build a robust automatic condition system one shall define a collection of parameters to be monitored. A change in any of these parameters would indicate a change in the structure. The scope of this work was not to build such a complete system but to test the system at a prototype and prove that it is effective. There is room for improvement, which will be achieved during the construction phase of the MSTs planned in the next years.

\bibliography{main} 

\begin{thebibliography}{10}

\bibitem{fermilat}
{F.Loparco, Fermi-LAT Collaboration}, ``Seven years of gamma-ray astrophysics
  with the fermi lat,'' {\em Nuclear and Particle Physics Proceedings}  (2016).

\bibitem{hessicrc}
{H. Prokoph for the H.E.S.S. Collaboration}, ``{The H.E.S.S. Experiment:
  Current Status and Future Prospects"},'' in [{\em International Cosmic Ray
  Conference}{\nolinebreak\hspace{0.1em}]},   {\bf 36} (2019).

\bibitem{magic}
{MAGIC Collaboration}, ``{The major upgrade of the MAGIC telescopes, Part I:
  The hardware improvements and the commissioning of the system},'' {\em
  Astroparticle Physics}~{\bf 72} (2016).

\bibitem{veritas}
{R.Mukherjee for the VERITAS Collaboration}, ``{Observing the energetic
  universe at very high energies with the VERITAS gamma ray observatory},''
  {\em Advances in Space Research}  (2018).

\bibitem{sciencewithcta}
{The Cherenkov Telescope Array Consortium},  [{\em Science with Cherenkov
  Telescope Array}{\nolinebreak\hspace{0.1em}]}, arXiv:1709.07997 (2018
  (eleventh edition)).

\bibitem{markusicrc}
{M. Garczarczyk et al. the CTA Consortium"}, ``Status of the medium-sized
  telescope for the cherenkov telescope array,'' in [{\em International Cosmic
  Ray Conference}{\nolinebreak\hspace{0.1em}]},   {\bf 34} (2015).

\bibitem{victoricrc}
{V. Barbosa Martins et al. for the MST-STR project of the CTA consortium}, ``{A
  Condition Monitoring Concept Studied at the MST Prototype for the Cherenkov
  Telescope Array},'' in [{\em International Cosmic Ray
  Conference}{\nolinebreak\hspace{0.1em}]},   {\bf 36} (2019).

\bibitem{geosig}
{GeoSIG}, ``{Geosig, Swiss to measure}.'' \url{www.geosig.com}.
\newblock (Accessed: 25 October 2020).

\bibitem{gantner}
{Gantner}, ``{Gantner Instruments}.'' \url{www.gantner-instruments.com/}.
\newblock (Accessed: 25 October 2020).

\bibitem{fdd}
{R. Brincker et al.}, ``Modal identification from ambient responses using
  frequency domain decomposition,'' {\em Proceedings of SPIE - The
  International Society for Optical Engineering 1}  (2000).

\bibitem{svibs}
{Svibs}, ``{Download Papers on Operational Modal Analysis}.''
  \url{www.svibs.com/literature}.
\newblock (Accessed: 25 October 2020).

\bibitem{damage}
{S. Gres et al.}, ``Hankel matrix normalization for robust damage detection,''
  in [{\em International Operational Modal Analysis
  Conference}{\nolinebreak\hspace{0.1em}]},   {\bf 8} (2019).

\end{thebibliography}
\bibliographystyle{spiebib} 

\end{document}